\documentclass[acmsmall, nonacm]{acmart}

\usepackage{hyperref}

\AtBeginDocument{%
  \providecommand\BibTeX{{%
    \normalfont B\kern-0.5em{\scshape i\kern-0.25em b}\kern-0.8em\TeX}}}

\begin{document}




\title{Social Learning and the Accuracy-Risk Trade-off in the Wisdom of the Crowd}

\author{Dhaval Adjodah}
\email{dval@mit.edu}
\author{Yan Leng}
\author{Shi Kai Chong}
\affiliation{%
  \institution{Media Lab, Massachusetts Institute of Technology}
}

\author{P. M. Krafft}
\affiliation{%
  \institution{Oxford Internet Institute, University of Oxford}
}

\author{Esteban Moro}
\affiliation{%
  \institution{Universidad Carlos III de Madrid, 28911 Madrid, Spain}
}

\author{Alex Pentland}
\affiliation{%
  \institution{Media Lab, Massachusetts Institute of Technology}
}

    \begin{abstract}
        How do we design and deploy crowdsourced prediction platforms for real-world applications where risk is an important dimension of prediction performance? 
        To answer this question, we conducted a large online Wisdom of the Crowd study where participants predicted the prices of real financial assets (e.g. S\&P 500). We observe a Pareto frontier between accuracy of prediction and risk, and find that this trade-off is mediated by social learning i.e. as social learning is increasingly leveraged, it leads to lower accuracy but also lower risk.
        We also observe that social learning leads to superior accuracy during one of our rounds that occurred during the high market uncertainty of the Brexit vote. 
        Our results have implications for the design of crowdsourced prediction platforms: for example, they suggest that the performance of the crowd should be more comprehensively characterized by using \textit{both} accuracy and risk (as is standard in financial and statistical forecasting), in contrast to prior work where risk of prediction has been overlooked.

    \end{abstract}

\maketitle

\section{Introduction}
    
    Crowdsourced prediction systems through approaches such as the Wisdom of the Crowd \cite{galton1907vox, golub2010naive} or Prediction Markets \cite{hanson2003combinatorial, arrow2008promise} have been successful in a range of domains such as predicting the reproducibility of scientific research \cite{dreber2015using}, estimating the caloric content of food items \cite{becker2017network}, and predicting stock market prices \cite{nofer2014crowds}. 
    
    From the perspective of platform designers who want to deploy such prediction systems at scale for real-life applications, one under-studied aspect of the performance of the crowd is the risk of the prediction of the crowd: over a collection of separate prediction tasks, it is important to measure the variance of the average accuracy --- in addition to the average accuracy of the crowd in predicting the realized outcome correctly --- as it provides a measure of the risk of the crowdsourced prediction system. This is because the crowd might be accurate on average (over these separate prediction tasks) but have a high variance in accuracy across tasks (i.e. for some prediction tasks, the accuracy might be quite poor), making it risky to employ this system for prediction. 
    This view is standard in statistical \cite{james2013introduction, domingos2000unified, geman1992neural} and financial \cite{joyce1970uncertainty, modigliani1997risk, ghysels2005there} forecasting applications, but it not commonly studied in the literature on the Wisdom of Crowds and many adjacent areas of collective intelligence research. 
    
    To demonstrate the importance of measuring risk in Wisdom of Crowds platforms, we deployed a large crowdsourced prediction task  in which we measure the accuracy and variance of the prediction system over 7 independent rounds of predicting financial asset prices (S\&P 500, gold and WTI Oil).
    
    \subsection{Contributions}
    \begin{itemize}
        \item We observe a Pareto frontier \cite{markowitz1952portfolio, gammerman2007hedging} between accuracy and risk: as the average accuracy of the crowd over the different prediction rounds increases, so does the variance in the crowd's predictive accuracy. We further observe that this trade-off is mediated by the amount of social learning---i.e., the extent to which users pay attention to each other's judgments. 
        \item We deployed one of our prediction tasks just before the Brexit vote during which there was a great deal of market uncertainty \cite{oehler2017brexit}, and we observe that during such uncertain times social learning leads to higher accuracy. 
        \item While modeling the belief update process of participants using Bayesian Models of Cognition \cite{Griffiths2006,Griffiths2008} to estimate their amount of social learning, we observe that our participants exhibit the attribute substitution heuristic of human decision-making \cite{Kahneman2014}, whereby a complicated problem is solved by approximating it with a simpler, less accurate model. We also observe people's preference to learn from social information rather than from non-social information. 
        \item We are releasing our large dataset\footnote{Data and code are available \href{https://osf.io/jxw45/?view_only=8b3e533334824f4d81efcc44dcb5d80d}{here}.} which is the first dataset, to the authors' knowledge, that records not only participants pre- and post-exposure predictions, but also both the social and non-social information they were exposed to in a large-scale social Wisdom of the Crowd domain.
    \end{itemize}

\section{Related Work}

    \subsection{Computer-supported cooperative work (CSCW)}
    In the present work we study how platform design mediates collective intelligence in the context of a Wisdom of the Crowd task.
        The study of collective intelligence---the ability of groups to come together to solve problems collectively---has long been a key area of research in computer-supported cooperative work (CSCW). Within the CSCW community, the interest in this area of research has centered on how to improve collaborative work through the lens of collective intelligence research. Building on early work of Malone, Grosz, and colleagues \cite{malone1990coordination,grosz1996collaborative}, CSCW researchers have studied factors that influence collective intelligence \cite{kim2017makes}, platforms for promoting collective intelligence \cite{de2012contested,amir2015care,malone2017putting}, frameworks for understanding collective intelligence \cite{grasso2012collective}, and phenomena of collective intelligence in digital settings (e.g., \cite{twyman2017black,jahani2018scamcoins}).
        
        More recently, a new strand of work has looked into how to deliver high-quality results for complex real-world applications: for example a system inspired by distributed computing infrastructures has allowed crowds to work with thousands of workers and tasks by accounting for human factors \cite{difallah2019deadline}, a hybrid system can overcome the issue of initial low-fidelity data \cite{garg20194x}, and new approaches has been presented that allow crowds to build datasets that approximate complex machine learning data distributions dynamically \cite{chung2019efficient}. 
        
        Similarly, we believe that in order to deploy Wisdom of the Crowd systems at scale --- for example, as in our task, the prediction of financial asset prices --- their performance must be more comprehensively characterized. 

    \subsection{Wisdom of the Crowd}
    
        One popular domain within the collective intelligence literature is the Wisdom of the Crowd \cite{galton1907vox, golub2010naive}, where participants (typically referred to as the `crowd') are asked to make predictions of a certain quantity, such as the future price of an asset on the stock market \cite{nofer2014crowds} or the caloric content of food items \cite{becker2017network}. It has been found that the central tendency of the crowd (such as the average, the median, or other aggregates) -- used as a measure of collective belief -- can be quite accurate \cite{galton1907vox, golub2010naive}, where accuracy is defined as the error between the crowd's aggregate prediction and the ``ground truth'' (here, the realized future price of the asset). 
        
        There is a rich literature aiming to optimize the accuracy of the crowd, such as by recalibrating predictions against systematic biases of individuals \cite{turner2014forecast}, selecting participants who are resistant to social influence \cite{madirolas2015improving}, rewiring the network topology of information-sharing between subjects \cite{almaatouq2020adaptive, becker2017network}, and optimally allocating tasks to individuals \cite{karger2014budget}. 
        
        Overall, it has been hypothesized that crowds can be highly accurate in aggregate because people's individual biases are typically not correlated \cite{nofer2014crowds}, and, therefore, cancel out on average. However, when participants in the crowd start sharing information, such as through social learning, their beliefs can become correlated and therefore degrades the accuracy of the crowd. In the next section, we discuss the impact of social learning on accuracy. 

    \subsection{Social Learning}

        One of the promising avenues for advancement in the CSCW field from the science of collective intelligence is the effect of social learning---the use of information about other people's decisions to make one's own --- on the collective performance of crowdsourcing systems such as the Wisdom of the Crowd. 
        
        Several threads of research in CSCW and adjacent areas examine the importance of social recommendation on engagement with web content (e.g., \cite{lerman2010information,lerman2010using,stoddard2015popularity,celis2016sequential}).
        From a design perspective, the relationship between social observation and collective intelligence  is especially interesting because crowdsourcing platforms can often be optimized to be more social \cite{lim2010stakesource, chen2015crowdsourcing}, even though research on collective prediction problems is divided on the effect of social learning on collective performance. 
        
        On one hand, prior work has shown that exposure to social information can lead to degraded performance in aggregate guesses \cite{Lorenz2011,muchnik2013social,turner2014forecast}: increasing the strength of social influence has been shown to increase inequality \cite{salganik2006experimental}, selecting the predictions of people who are \textit{resistant} to social influence has been shown to have improved collective accuracy \cite{madirolas2015improving}, the influence of influential peers has been theoretically shown to prevent the group from converging on the true estimate \cite{turner2014forecast}, and exposure to the confidence levels of others has been shown to influence people to change their predictions for the worse \cite{Moussaid2013}.
        
        On the other hand, social learning has also been shown to lead to groups outperforming their best individuals when they work separately \cite{almaatouq2020adaptive}, a collective intelligence factor has been shown to predict team performance better than the maximum intelligence of members of the team \cite{woolley2010evidence}, and human-inspired social communication between agents has been shown to improve collective performance in optimization algorithms \cite{lazer2007network, adjodahNetES}.  
        
        Therefore, the role of social learning in collective performance is still being understood, but the question of how social learning impacts collective intelligence has great potential to impact our understanding of platform and interface design in CSCW settings.
    
    \subsection{Risk}
        The prior work mentioned above in both the social learning and the Wisdom of the Crowd literatures have focused on maximizing the average accuracy of groups with little regard to the variance (risk) of the predictions. It has been proven theoretically \cite{james2013introduction,domingos2000unified} and observed in a variety of statistical applications \cite{geman1992neural, gagliardi2011instance} that there is a fundamental trade-off between accuracy and risk. This means that, for any prediction system, risk will be ever-present and that maximizing accuracy will lead to increasing risks, i.e. the performance of the system will always exist within a pre-defined Pareto frontier \cite{markowitz1952portfolio, gammerman2007hedging}. 
        
        In practice, treating risk and accuracy as equally important for prediction is standard in statistical \cite{james2013introduction, domingos2000unified, geman1992neural} and financial \cite{joyce1970uncertainty, modigliani1997risk, ghysels2005there} forecasting applications and literature because it allows system designers to carefully calibrate their strategies to risk --- for example to hedge for probable losses \cite{chavez2006quantitative, asmussen2006improved, shevchenko2006structural, chapelle2008practical, cruz2002modeling}. Furthermore, in a meta-study of 105 forecasting papers, 102 of them support prioritizing for lower risk \cite{armstrong2015golden}.
        
        At the individual level, there is strong evidence that people preferentially optimize for risk instead of accuracy in a variety of domains \cite{holt2002risk}: cognitively, people have been observed to manifest decision heuristics \cite{kahneman2013prospect} to be conservative in the face of uncertainty \cite{hintze2015risk,zhang2014origin}. For example, rice farmers have been observed not to adopt significant harvest improvement technology because of the risk of it failing once and causing significant family ruin \cite{binswanger1983risk}. Evolutionarily, risk aversion has been shown to emerge when rare events have a large impact on individual fitness \cite{hintze2015risk}. Theoretically, when considering more realistic decision-making scenarios, such as repeated trials (as opposed to one-time bets), it has been shown that accounting for risk is critical to understanding the dynamics of how people make decisions \cite{peters2019ergodicity}. 
        
        
        Given the importance of risk both at the individual level and as an important metric in a range of forecasting applications, it is important to study it within the context of crowdsourced prediction platforms. 
        
        
        

     \begin{figure*}[t]
        \centering
        \includegraphics[width=.9\textwidth]{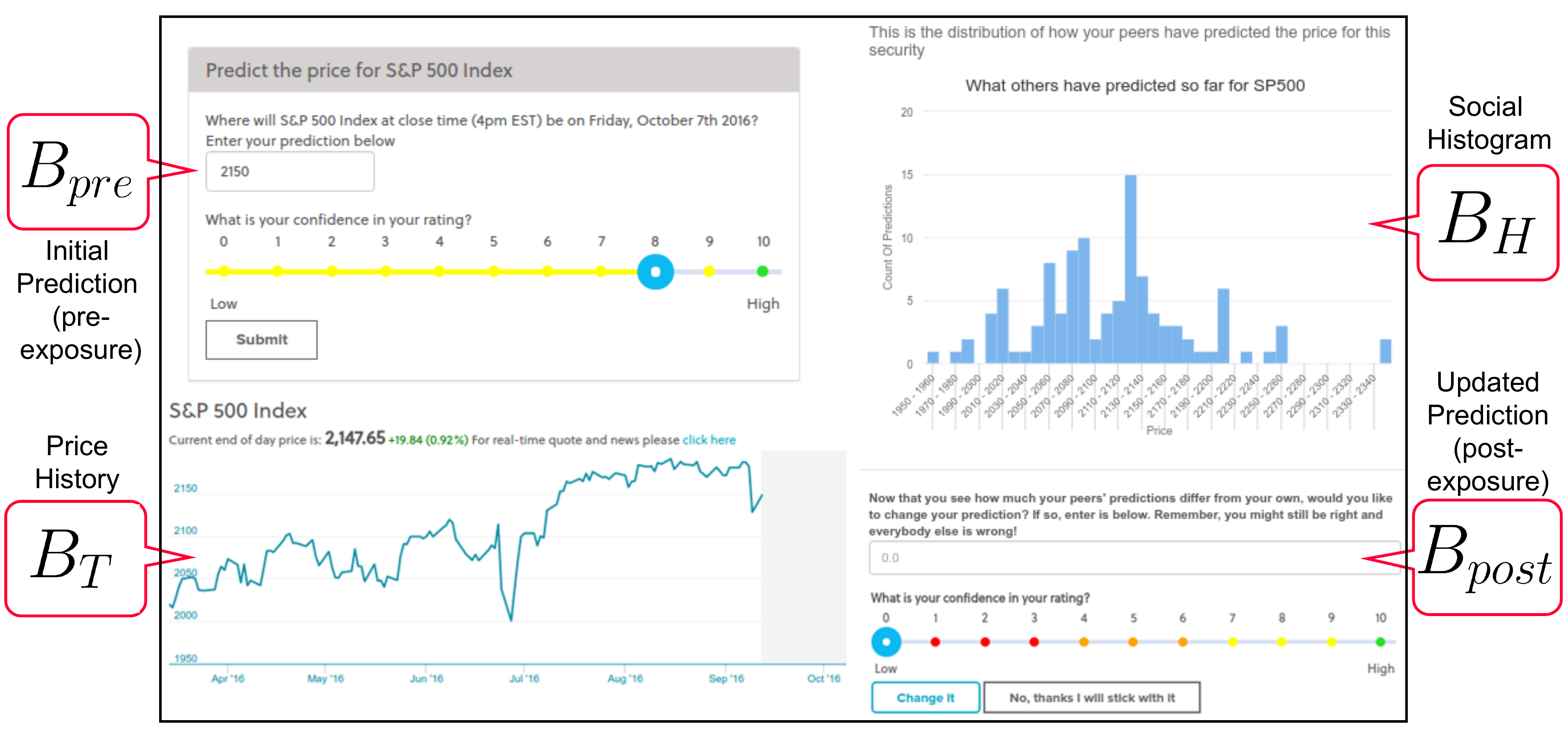}
        \caption{An annotated screenshot of how data was collected: the pre-exposure prediction $B_{pre}$ is shown first, followed by the social histogram ${B_{H}}$ and the price history ${B_{T}}$. Finally, the updated prediction $B_{post}$ is collected. The ground truth of the asset's final closing price will be $V$ (not shown here, realized at the end of the round).
        }
        \label{fig:screenshot_nice}
    \end{figure*}
    
\section{Experimental Design}
    
    
    The main goal of our study is to investigate if risk is an important dimension of crowdsourced prediction platforms that is required to fully characterize the performance of the crowd, and whether social learning affects risk. To do so, we hypothesize that a Pareto frontier exists between risk and accuracy --- i.e. that there is a trade-off between risk and accuracy of prediction across several prediction rounds --- and that social leaning impacts this trade-off. 
    
    To test this hypothesis, we need a dataset with the following requirements:
    \begin{itemize}
        \item Predictions made of complex and difficult to predict phenomena so that our results are applicable to the real-world platform applications being studied within the CSCW community. 
        \item A standard against which we can compare our dataset to judge its external validity.         
        \item A large number of predictions for statistical significance, with both pre- and post-exposure predictions collected. 
        \item The exact social and non-social information each user was exposed to after their initial pre-exposure prediction so that we can later model how different types of information influenced them in updating their belief into their post-exposure prediction. 
        \item Predictions over many separate independent prediction rounds. 
        \item At least one prediction round that occurred during a period of high uncertainty to understand if our findings change in abnormal settings. 
    \end{itemize}
    
    Given the above requirements, we designed the experimental procedure detailed below: we recruited a total of 2,037 participants over seven prediction rounds to predict the future prices of financial assets (the S\&P 500, WTI Oil, and gold prices) during seven separate consecutive 3-week rounds over the span of 6 months, resulting in 9,268 predictions (i.e. 4,634 prediction pairs/sets). We focused on predicting financial prices as doing so is a hard prediction problem \cite{campbell1988stock,fama1995random}. Our participants were mid-career financial professionals with years of financial experience. Our participants consented to their data being used in this study and we obtained prior IRB approval.
    
    \subsection{Data Collection}
    As shown in the screenshot of the user interface in Fig.  \ref{fig:screenshot_nice}, we designed the data collection process as follows: every time a user makes a prediction of an asset's future price through our platform, the following prediction set comprising $B_{pre}, {B_{H}}, {B_{T}}$ and $B_{post}$ is collected:
    \begin{itemize}
    \item A ``pre-exposure'' belief prediction $B_{pre}$, which is independent of any social information. For example, a user might show-up on the platform and predict that the closing price of the S\&P 500 to be \$2,001 on June 24$^{th}$, 2016. 
    \item The predictions ${B_{H}}$ within the social information histogram shown to each user after each initial prediction. Additionally, we display a 6-month time-series of the asset's price ${B_{T}}$ up to this point.

    \item The revised ``post-exposure'' prediction $B_{post}$. For example, after seeing the social histogram and asset price history, a user might update their belief to \$2,201. Since the real price (the ground truth $V$) ended up being \$2,037.41, this user became more accurate after information exposure (they went from \$2,001 to \$2,201).
    \end{itemize}

    We ensure that the ``pre-exposure'' prediction is made before any social information is shown. We present a unique histogram for every new prediction (as it is built using past predictions up to this point), as well as a unique price history time series (as it shows the 6-month price data up to the time of prediction). We require all participants to make a post-exposure prediction.  
    
     During each round, participants made a prediction of the same asset's closing price for the same final day of the round. We use the round's last day's closing market price as our measure of ground truth. We carefully instrumented the social and non-social information that our participants were exposed to, and collected their predictions before and after exposure to this information. We also deployed one of our rounds during a high uncertainty period to understand if variance reduction strategies allows the crowd to be resistant to risk. 
     
     Whenever we predict a final closing price, we only use user prediction data up to the week before the day of prediction (i.e., we don't use any data during the last week of the round) so that our predictions are not too easy. One of our rounds of prediction happened to end the day of the Brexit vote, which means that we have prediction data during a particularly volatile market period \cite{oehler2017brexit}. We chose the start and end dates of each round so that the expiry dates of the asset's underlying \textit{futures} would not affect the price of both the asset and its futures. Financial data (asset and futures prices) is obtained through Barchart.com's API.

    \subsection{External Validity of Data Collected}
        
        As shown in Table \ref{tab:table_assets}, our participants are collectively quite accurate --- in agreement with past Wisdom of the Crowd studies \cite{nofer2014crowds, golub2010naive} --- indicating that their predictions are being thoughtfully elicited: we observe that the crowd is generally doing more than just linear extrapolation (we test a model where we simply extrapolate prices in time using a static slope) as the error of such a model is higher. Additionally, we observe that the crowd's mean prediction error\footnote{Higher relative errors in round 2 are an artifact of the fact that a few dollars' error on the lower price of WTI Oil seems like a higher error compared to same absolute error on the higher prices of the other assets (about \$45 per share for WTI Oil compared to \$2100 for the S\&P 500 and \$1300 for gold prices).} is much less than the overall price change of the asset for the 3-week prediction period.
        
        Interestingly, the crowd's collective prediction over each round tracks (and sometimes outperforms) the futures of each asset being predicted (we calculate the futures error as the difference between the futures price and the asset price). Because futures prices are commonly used a measure of the global market's prediction of the price asset \cite{alquist2010we, french1986detecting}, the fact that the crowd's performance is on-par with the futures prices indicates that our dataset is externally valid.
        

        

        \begin{table}[h]
        \centering
        \resizebox{.8\columnwidth}{!}{
        \begin{tabular}{|l|l|l|l|l|l|l|l|}
        \cline{2-8} 
        \multicolumn{1}{l|}{} & \multicolumn{7}{c|}{\textbf{Prediction Round}}\tabularnewline
        \cline{2-8} 
        \multicolumn{1}{l|}{} & \textbf{1}  & \textbf{2}  & \textbf{3}  & \textbf{4}  & \textbf{5}  & \textbf{6}  & \textbf{7} \tabularnewline
        \hline 
        \textbf{Asset}  & \textbf{SP500}  & \textbf{WTI Oil}  & \textbf{Gold}  & \textbf{SP500}  & \textbf{SP500}  & \textbf{SP500}  & \textbf{SP500} \tabularnewline
        \hline 
        \textbf{Grounth Truth (\$)} & 2037.41  & 45.95  & 1335.80  & 2153.74  & 2126.41  & 2191.95  & 2262.53 \tabularnewline
        \hline 
        \textbf{Num. Prediction Sets}  & 284  & 207  & 134  & 1174  & 925  & 1441  & 469 \tabularnewline
        \hline 
        \textbf{Price Change (\%)}  & 4.01  & 11.03  & 3.63  & 1.77  & 1.75  & 2.24  & 3.56 \tabularnewline
        \hline 
        \textbf{Linear Extrapolation Error} \textbf{(\%)} & 6.66  & 16.4  & 1.26  & 1.62  & 2.75  & 0.75  & 3.10 \tabularnewline
        \hline 
        \textbf{Crowd Mean Error} \textbf{(\%)} & 2.22  & 4.95  & 0.46  & 0.84  & 0.58  & 3.20  & 2.40 \tabularnewline
        \hline 
        \textbf{Futures Mean Error} \textbf{(\%)} & 2.03  & 3.05  & 0.94  & 0.38  & 0.40  & 0.48  & 1.50 \tabularnewline
        \hline 
        \end{tabular} 
        }
        \caption{Summary of data collected. Our crowd is accurate, and sometimes even outperforms the futures underlying the asset. Predictions made by participants are more accurate than simple linear extrapolation.}
        \label{tab:table_assets}
        \end{table}

\subsection{Brexit Data}
\begin{figure}
            \centering
            \includegraphics[width=0.5\textwidth]{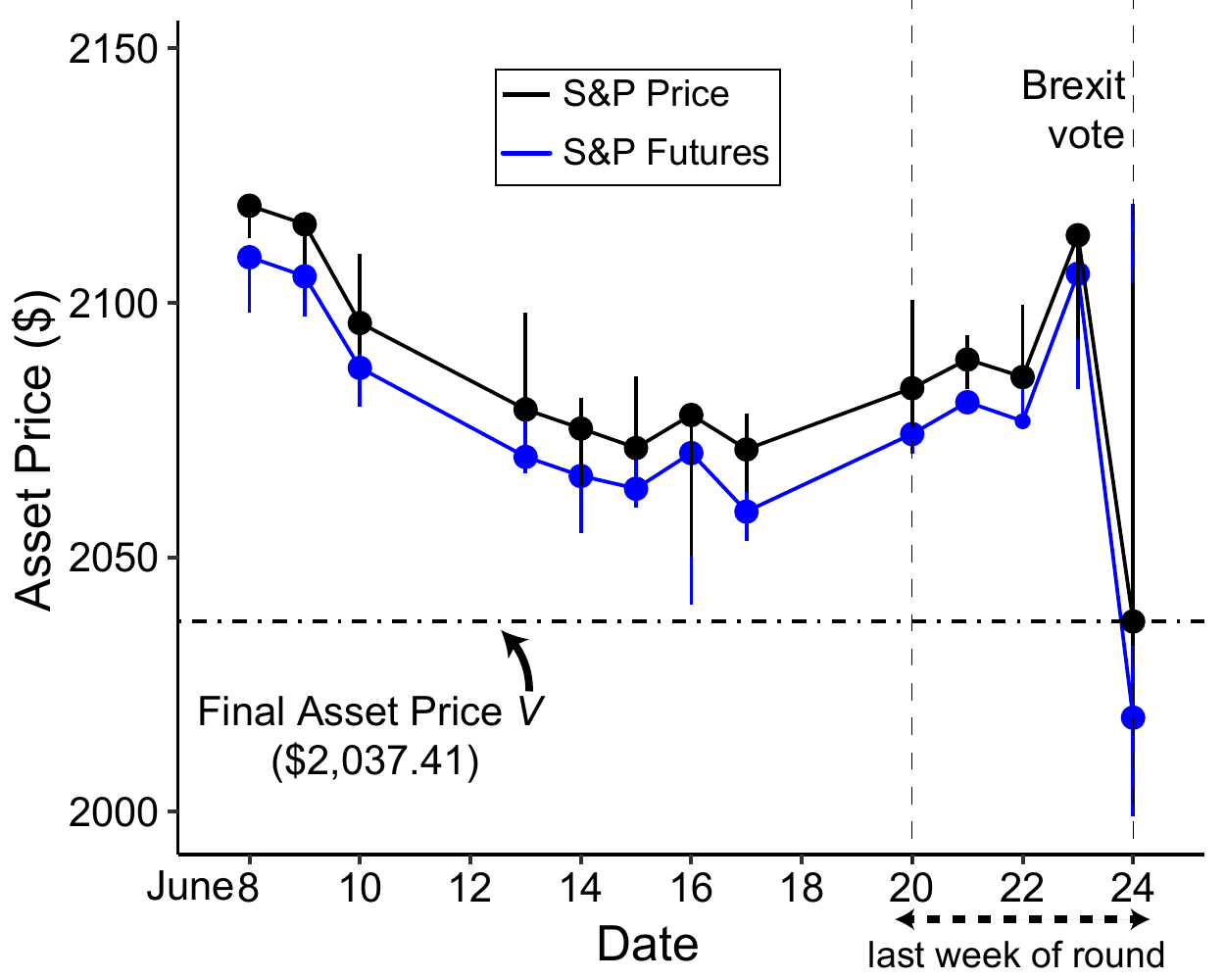}
            \caption{
            The close, low, and high price of the asset and its underlying futures are shown as candlestick plots. The asset and futures overestimated the price and then crashed during the last week.
            }
            \label{fig:brexit_accuracy_line}
            \end{figure}

    We deployed one of our experiments right before the Brexit vote during which there was a lot of market uncertainty \cite{oehler2017brexit}: the prediction round starting on June 1$^{st}$ 2016 ended on June 24$^{th}$ 2016, the day of the Brexit vote, and participants were predicting the price of the S\&P 500, an asset sensitive to global events \cite{deshpande2020brexit, cox2018political}. We collected 284 prediction sets during the first 2 weeks of the round, and 52 sets in the last week during which the global financial market first overestimated then underestimated the final price of the S\&P 500 asset leading to a 3.7\% crash, as shown in the candlestick plot in Fig. \ref{fig:brexit_accuracy_line}.

\section{Methods}

    Our hypothesis is that a Pareto frontier exists between risk and accuracy --- i.e. that there is a trade-off between risk and accuracy of prediction across several prediction rounds --- and that social learning impacts this trade-off. 
    
    In order to test this hypothesis, we need to select subsets of predictions based on how much social learning impacted the way these predictions were made. We can then calculate the risk and accuracy of these subsets and study the impact of social learning on these two aspects of performance. Given that participants were exposed to \textit{both} social and non-social learning, we cannot directly separate predictions based on whether they were updated through social learning or not. We therefore need a way to \textit{estimate} how much social learning was used by participants for each prediction, and then a way to select subsets of prediction based on their amount of social learning. Using these subsets, we then want to measure how much social learning impacted accuracy in this subset \textit{compared} to the full set of predictions (namely, the Wisdom of the Crowd) - we calculate not just the accuracy of a subset but its improvement relative to the original Wisdom of the Crowd (average of the full set of predictions).  
    
    From the perspective of platform designers who want to be able to select predictions based on required levels of accuracy or risk (e.g. to fit a certain portfolio of risk), it is important to measure improvement of subsets relative to the full collection of predictions. This is because, currently, platform designers only have access to one global measure of risk and accuracy --- that of the whole set of predictions (when there is no subset filtering). To demonstrate that selecting subsets of predictions can lead to significant \textit{improvements} in accuracy and risk, we therefore need to calculate these improvements. 
    
    For our results to support our hypothesis, we need to find a statistically significant trade-off between risk and accuracy over subsets with varying amounts of social learning. 
    
    The structure of this section is as follows:
    \begin{itemize}
        \item In section \ref{method:beliefupdate}, we describe how we model individual belief update ---  how a participant updates their prediction from their pre- to post-exposure prediction --- using either numerical Monte Carlo methods or approximate methods based on prior work by \cite{Griffiths2006, Griffiths2008}.   
        \item In section \ref{method:modelerror}, we detail how we evaluate the error (residual) between the \textit{modeled} updated belief and the actual updated belief of a participants. This error will allow us to introduce a parameter which allows us to estimate the relative amount of social learning for each prediction set. Creating such parameters is standard in the Wisdom of the Crowd literature. \cite{soll2009strategies, kerckhove2016modelling}. 
        \item In section \ref{method:subsetting}, we describe our methodology for subsetting predictions based on their estimated amount of social learning. Subsetting predictions based on estimated quantities is common \cite{madirolas2015improving}.  
        \item In section \ref{method:evaluationsubset}, we explain how the accuracy and risk of subsets are measured, and how they are used to create a Pareto curve \cite{markowitz1952portfolio}. 
    \end{itemize}


    
    
\subsection{Modeling Belief Updates}
\label{method:beliefupdate}
    

        Using formalism inspired by Bayesian models of cognition \cite{Griffiths2008}, we can model the 4,634 prediction sets collected over many rounds, at a high level, as a Bayesian update.
        To use this formalism, we need to select a prior distribution for each individual's belief, a likelihood (evidence) distribution, and a way to use them to compute the posterior (updated belief) distribution. Here, we describe modeling at a high level, and describe more thoroughly our two modeling approaches in the next sections. 
        

        As we are interested in how individuals update their belief regarding the asset's future price (ground truth) $V$ based on the information we expose them to, the choice of the prior distribution is straightforward: $P_{prior}(V)\approx P(B_{pre})$, the distribution over an individual's pre-exposure belief of $V$. 
        There are two main likelihood (evidence) distributions participants employ: the assets' price history $B_{T}$ participants are shown, giving us $P_{likelihood}(V) \approx P(B_{T})$, or analogously, the social histogram $B_{H}$, giving us $P_{likelihood}(V) \approx P(B_{H})$. 
        
        Given the prior and likelihood, the \textit{modeled} posterior prediction $P_{posterior}(V)$, can, therefore, be approximated as $P_{posterior}(V) \propto P(B_{H})\cdot P(B_{pre})$ in the case of the social histogram, and $P_{posterior}(V) \propto P(B_{T})\cdot P(B_{pre})$ when participants learn from the past price history. When using the social histogram, we can simply bin the prices shown to a participant and obtain a distribution over prices. When using price history ${B_{T}}$, the time-series of prices needs to first be transformed into a `rates' histogram as is standard in financial technical analysis \cite{park2007we, neftci1991naive}. To do so, the daily rate in price change is calculated, and this histogram of price change per day is used to extrapolate and predict asset prices. Specifically, a daily rate, $r_t$, of asset price change is calculated for each day during the 6-month interval that a user is shown, $r_t = \frac{B_{t} - B_{t-1}}{B_{t}}$. These rates are then used to create a histogram of prices similar to when using the social histogram.
        
        

        We do not make any other assumptions in terms of what data to use to approximate the likelihood and prior distributions. Given these distributions, the question is then how to compute the posterior (updated) belief of an individual.
        
        Although the space of possible distributions and posterior computation approaches is very large, we focus here on using two simple, interpretable, and theoretically-motivated approaches from prior work \cite{Griffiths2006}, namely using Gaussian (normal) distributions to approximate priors and likelihoods, and using a Monte Carlo numerical sampling approach to calculate the posterior from the actual distributions of prices that participants were exposed to. We leave to future work the exploration of richer distributions and approaches to modeling belief update as it is beyond the scope of this study.

    \subsubsection{Approximate Approach:}        
    \label{method:parametric} 
        In this approach, we assume both the prior and likelihood to be normally distributed such that the \textit{modeled} posterior --- which models the belief update of an individual and therefore allows us to predict their belief after exposure to information --- is also normally distributed. When using the social histogram ${B_{T}}$ as evidence, the posterior is $P_{posterior}(V) = (B_{pre} + \overline{{B_{H}}})/2$. We call this model \texttt{GaussianSocial}\footnote{We can sum the scalar $B_{pre}$ to the average of ${B_{T}}$, $\overline{{B_{T}}}$, as the average is also a scalar.}w.
        When modeling belief update from the price history ${B_{T}}$ we obtain \texttt{GaussianPrice}, where $P_{posterior}(V) = (B_{pre} + \overline{{B_{T}}})/2$. We include the derivation of these models in the supplementary. 
    
    \subsubsection{Numerical Approach:}
    \label{method:montecarlo} 
        Instead of using an approximated distribution for the likelihood, following the formalism of \cite{Griffiths2006}, we can use a numerical approach by binning the likelihood distributions to estimate the posterior distribution using Monte Carlo methods. Because we do not have access to the distribution of the prior belief of each individual (as we only have an individual point estimate for each prediction set), we still have to approximate the prior. We model the prior to be Gaussian, with the mean set as the pre-exposure prediction of an individual, $B_{pre}$, and the standard deviation set as the standard deviation of the social histogram ${B_{H}}$ or the standard deviation of the price history ${B_{T}}$, depending on which likelihood distribution was used. 
        
        Specifically, we calculate the posterior distribution $P_{posterior} (b)$ of an individual's post-exposure prediction $b$ in the following way: 
        let $b_j$ be a unique value in $\mathbf{B_{H}}$, and $P_{B_H}(b_h)$ be the probability density of $b_h$ in $\mathbf{B_H}$. Let $P_{prior}(b)$ be the density of $b$ in the parametrized prior distribution. The posterior distribution for the numerical model is defined as 
        $
            P_{posterior} (b) = \frac{P_{B_{H}}(b)  \times  P_{prior}(b) }{ \sum_{{b_j} \in {B_{H}}} P_{B_H}(b_j)  \times  P_{prior}(b_j) }
        $ 
        when using the social information ${B_{H}}$. 
        After computing this posterior distribution using rejection sampling \cite{gilks1995adaptive} --- our data and distribution are small enough that rejection sampling was fast enough ---, we use the mean of the distribution as the \textit{modeled} updated belief of a participant.  
        

    \subsection{Evaluating Model Error}
    \label{method:modelerror}
        For all models, we compute the relative residual error between the model's prediction of the posterior ($ \mu_{\sim P_{posterior}(V)}$) and the actual post-exposure prediction ($B_{post}$) as: $ (\mu_{\sim P_{posterior}(V)} - B_{post})/B_{post}$. For the approximate approach, $ \mu_{\sim P_{posterior}(V)}$ is simply the mean of the normal distribution representing the posterior, while in the numerical approach, the mean is estimated through averaging over all bins of the empirical distribution (the distribution is small enough that sampling was not needed). 

        For all models, the 95\% confidence intervals are calculated as follows: we assume the data follows Student's t-distribution since the variance of the true distribution is unknown and, therefore,  we estimate it from the sample data. Let $s_e$ be the estimated standard error of the sample mean and $t_e$ be the t-value for the 95\% confidence interval desired, which can be computed via inverse t-distribution. The lower and upper limits for the 95\% confidence interval are $[\mu_e - t_e s_e, \mu_e + t_e s_e]$, where $\mu_e$ is the estimated sample mean.

    \subsection{Subsetting Predictions}
    \label{method:subsetting}
    
        Using these models, we can \textit{estimate} which information source --- social information or price history --- each participant used to update their belief by comparing the residual errors of models using either social information or price history as likelihood. This will allow us to select subsets of prediction based on whether they were more likely updated using social or non-social information. 
        
\begin{figure}[t]
                \centering
                \includegraphics[width=.7\columnwidth]{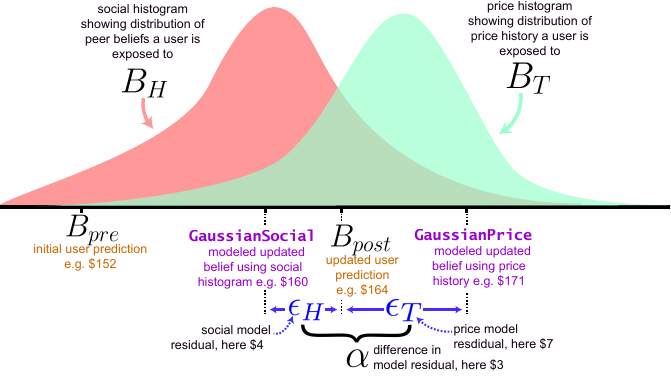}
                \caption{An example belief update: for each prediction set, a user updates their belief from the pre-exposure prediction $B_{pre}$ to the updated prediction $B_{post}$ by either learning from the social histogram ${B_{H}}$ and/or the price history ${B_{T}}$. $\epsilon_{H}$ is the residual between the \emph{modeled} updated prediction \texttt{GaussianSocial} and the participant's updated prediction $B_{post}$; $\epsilon_{T}$ is the residual between \texttt{GaussianPrice} and $B_{post}$. $\alpha$ is the difference between $\epsilon_{T}$ and $\epsilon_{H}$.
                }
                \label{fig:two_residuals_illustration}
        \end{figure}

        Our approach is illustrated in Fig. \ref{fig:two_residuals_illustration}. 
        This approach of estimating characteristics of how predictions are revised is standard in the Wisdom of the Crowd literature. For example, prior work has estimated resistance to social influence \cite{madirolas2015improving} and influenceability in revising judgements after seeing the opinion of others \cite{kerckhove2016modelling, soll2009strategies}, and used them to improve collective performance. 
        
        Although we explored many models of belief update (as detailed in Result section \ref{result:beliefupdate}), we choose to focus on the \texttt{GaussianSocial} and \texttt{GaussianPrice} models (which assume the prior and likelihood to be Gaussian) due to their superior modeling accuracy, and because they are simple, interpretable, and theoretically-motivated models from prior work \cite{Griffiths2006}. We leave for future work the interesting question of searching the large space of parametric and non-parametric models and distributions to best predict people's belief update process. 
                    
        Therefore, using \texttt{GaussianSocial} and \texttt{GaussianPrice}, we calculate a residual $\epsilon_{H}$ for when learning from social information ${B_{H}}$ and a residual $\epsilon_{T}$ when learning from the price history ${B_{T}}$, as $\epsilon_{H} = \frac{|\texttt{GaussianSocial} - B_{post}|}{B_{post}} $ and $\epsilon_{T} = \frac{|\texttt{GaussianPrice} - B_{post}|}{B_{post}}$ respectively. We define $\alpha = \epsilon_{T} - \epsilon_{H} $, and we use it to measure how likely a participant used each source of information to update their prediction. For example, for a prediction set $[B_{pre}, {B_{H}}, {B_{T}}, B_{post}]$ if $ \alpha > 0$ (i.e., $\epsilon_{T} > \epsilon_{H}$), this means that this prediction set is better modeled using the social histogram of peer's belief ${B_{H}}$ instead of the price history ${B_{T}}$.
        
        Using $\alpha$, which we re-scale to be in the interval [-1,1] for each round, we can select a subset $S_{\alpha_{s}}$ of the prediction sets such that the $\alpha$ of these prediction sets lie in the range $0\leq \alpha < \alpha_{s}$ (or $\alpha_{s} < \alpha \leq 0$ when $\alpha_{s} < 0$), where $\alpha_{s}$ is the one-sided boundary we will vary in order to measure how much more likely a participant updated their belief from the social information instead of the price history. For example, the higher $\alpha_{s}$ is, the more likely a prediction set is better modeled using the social histogram of peer's belief ${B_{H}}$ instead of the price history ${B_{T}}$.


    \subsection{Evaluating Improvement of Subsets}
    \label{method:evaluationsubset}
    
        Each prediction set is now associated with a measure of the relative amount of social vs non-social learning, the parameter $\alpha$. 
        To subset predictions based on $\alpha$, we bin the $\alpha$'s from all 4,634 prediction sets into 15 groups of equal size,
        and compute the improvement in prediction error of each subset $S_{\alpha_{s}}$ and its variance compared to when all the crowd's predictions are used  (i.e. compared to \textit{the} Wisdom of the Crowd).
        To select a subset $S_{\alpha_{s}}$ of the prediction sets, we select them based on whether their the $\alpha$ of these prediction sets lie in the range $0\leq \alpha < \alpha_{s}$ (or $\alpha_{s} < \alpha \leq 0$ when $\alpha_{s} < 0$). 
        In order to measure the \textbf{improvement} in accuracy after exposure to information, we select only the post-exposure (updated) predictions $\pi^{post}_{j}$ within the subset of predictions $S_{\alpha_{s}}$, calculate the average prediction of this subset $\mathop{\mathbb{E}}_{j \sim S_{\alpha_{s}}}[\pi^{post}_{j}]$, and then the (absolute) error of this average within each round $i$, with respect to the ground truth $V$, as $\frac{\mathop{| \mathbb{E}}_{j \sim S_{\alpha_{s}}}[\pi^{post}_{j}] -V |}{V}$
        
        
        Similarly, we then calculate the error of the whole crowd's post-exposure predictions ${S_{all}}$, through the same computation as above, using $-1 \leq \alpha \leq 1$ for the set ${S_{all}}$ instead of $S_{\alpha_{s}}$. We define improvement $I^{S_{\alpha_{s}}}$ as the absolute difference between these two errors, as it measures the improvement in accuracy of using a subset $S_{\alpha_{s}}$ over using the full set of predictions --- the Wisdom of the Crowd --- within ${S_{all}}$. Critically, this will allow us to measure if exposure to varying degrees of social vs non-social information improves or worsen the performance of the crowd, and is an important metric for platform designers looking to improve the performance of a crowdsourced prediction system. 
        
        Note that the improvement defined so far is for each bootstrap $b$ for each round $i$, and is more clearly denoted as $I^{S_{\alpha_{s}}}_{i,b}$. Our reported value of improvement (the one in Fig. \ref{fig:Pareto}) is over 100 random bootstraps with replacement and is thereby calculated as such: over all rounds $i$, we first calculate the average improvement, $\displaystyle \mathop{\mathbb{E}}_{i} [I^{S_{\alpha_{s}}}_{b,i}]$ for each bootstrap $b$, and then, over all bootstraps, we calculate  $\displaystyle \mathop{\mathbb{E}}_{b} [\displaystyle \mathop{\mathbb{E}}_{i} [I^{S_{\alpha_{s}}}_{b,i}]]$. We use boostrapping \cite{efron1992bootstrap} in order to have a more robust estimate of the average accuracy and its variance (described in the next section). 
        

        In a Pareto curve we are interested in not only measuring the average accuracy described above but also the risk of the crowd in predicting the wrong collective prediction \textbf{over the different rounds}. This is because we are interested in estimating the spread of the distribution of accuracy --- risk --- of the Wisdom of the Crowd over all prediction rounds.
        Therefore, in Fig. \ref{fig:Pareto}, we first calculate, for each bootstrap $b$ the standard deviation (our measure of risk) across the seven rounds $i$ of prediction, i.e. $\sqrt{\displaystyle \mathop{\mathbb{E}}_{i} [( I^{S_{\alpha_{s}}}_{b,i} - \displaystyle \mathop{\mathbb{E}}_{i} [I^{S_{\alpha_{s}}}_{b,i}])^2]}$. We then compute the average of this risk over 100 bootstraps and report this value. We use standard deviation instead of variance as it is the more popular measure of risk in practice \cite{modigliani1997risk}.
        
        

    \subsection{Summary of Methods}
        We first model the belief update of individuals after they have been exposed to either the social information or the price history, using either an approximate approach or a numerical approach. We observe the residual (error) between the \textit{modeled} and the real updated belief of a participant using these various models and find that the \texttt{GaussianSocial} and \texttt{GaussianPrice} models (which assume the prior and likelihood to be Gaussian) outperform all other models (as detailed in Result section \ref{result:beliefupdate}). Using these models, we compute a parameter $\alpha$ which allows us to estimate whether each prediction set was more likely updated using social information or the price history. Using $\alpha$, we can select subsets of predictions that were more likely to have been made using one of the information sources. The accuracy this subset can then be compared to the full set of predictions (referred to as the Wisdom of the Crowd) to calculate the improvement of this subset. Similarly, we calculate the variance (risk) of this improvement.

    
    

\section{Results}

    \subsection{Accuracy-Risk Trade-off}
        
        
        \begin{figure}
        \centering
        \includegraphics[width=.5\columnwidth]{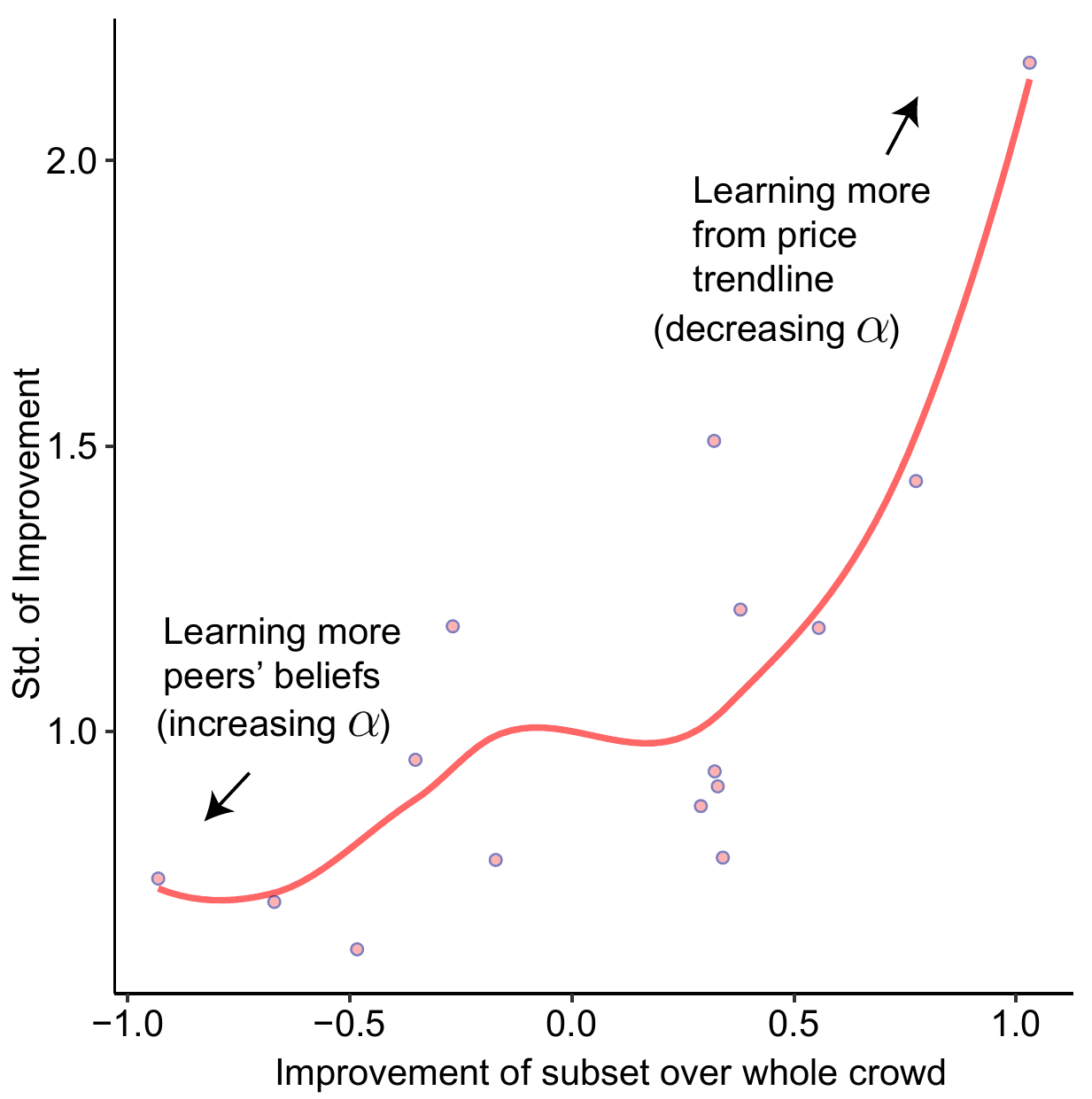}
        \caption{In this Pareto curve, we plot the improvement of each subset vs. the standard deviation in improvement within this subset. We see a risk-return trade-off: predictions made with price history are more accurate, but with higher risk (standard deviation). Smoothed curved is generated using LOESS \cite{cleveland1988locally}. 
        }
        \label{fig:Pareto}
        \end{figure}%
        
        Using a Pareto curve \cite{markowitz1952portfolio}, we compare the improvement in prediction accuracy 
        and risk of each subset $S_{\alpha_{s}}$. 
        As shown in the Pareto plot in Fig. \ref{fig:Pareto}, we observe that although people who learn more from price history are more accurate, there is increased variance---and therefore risk---in their predictions. This suggests that there is a risk-return trade-off between learning from one's peers versus looking at the price history: as social learning is increasingly leveraged, it leads to lower accuracy but also lower risk (replicating prior findings that exposure to social information decreases the variance of the crowd \cite{Lorenz2011}). Note that the social histogram is quite often non-unimodal (as detailed in the next result section \ref{result:beliefupdate}), which means that participants are intentionally collapsing multiple distribution modes to decrease variance.
        
        Such a Pareto trade-off between risk and accuracy is common in financial forecasting \cite{modigliani1997risk, ghysels2005there} and statistical prediction \cite{james2013introduction,domingos2000unified, geman1992neural,gagliardi2011instance}, but has not been typically observed in the literature on the Wisdom of Crowds. This has implications for the design of crowdsourced prediction platforms as described in the Discussion section \ref{discussion:design}.
                
        

    \subsection{Performance under High Uncertainty}

        An additional result of our study is the investigation of the crowd's performance during high uncertainty using the data from the prediction round that happened during the Brexit vote. We ran the analysis described earlier, but only for predictions made during the last week. Note that in all previous results, we took care not to use the last week of data to calculate collective accuracy so that prediction was not too easy, but we do so here as the high uncertainty makes prediction quite hard. This last week of data that we use is a disjoint subset of data from the data we previously used.

        \begin{figure}
        \centering
        \includegraphics[width=.5\columnwidth]{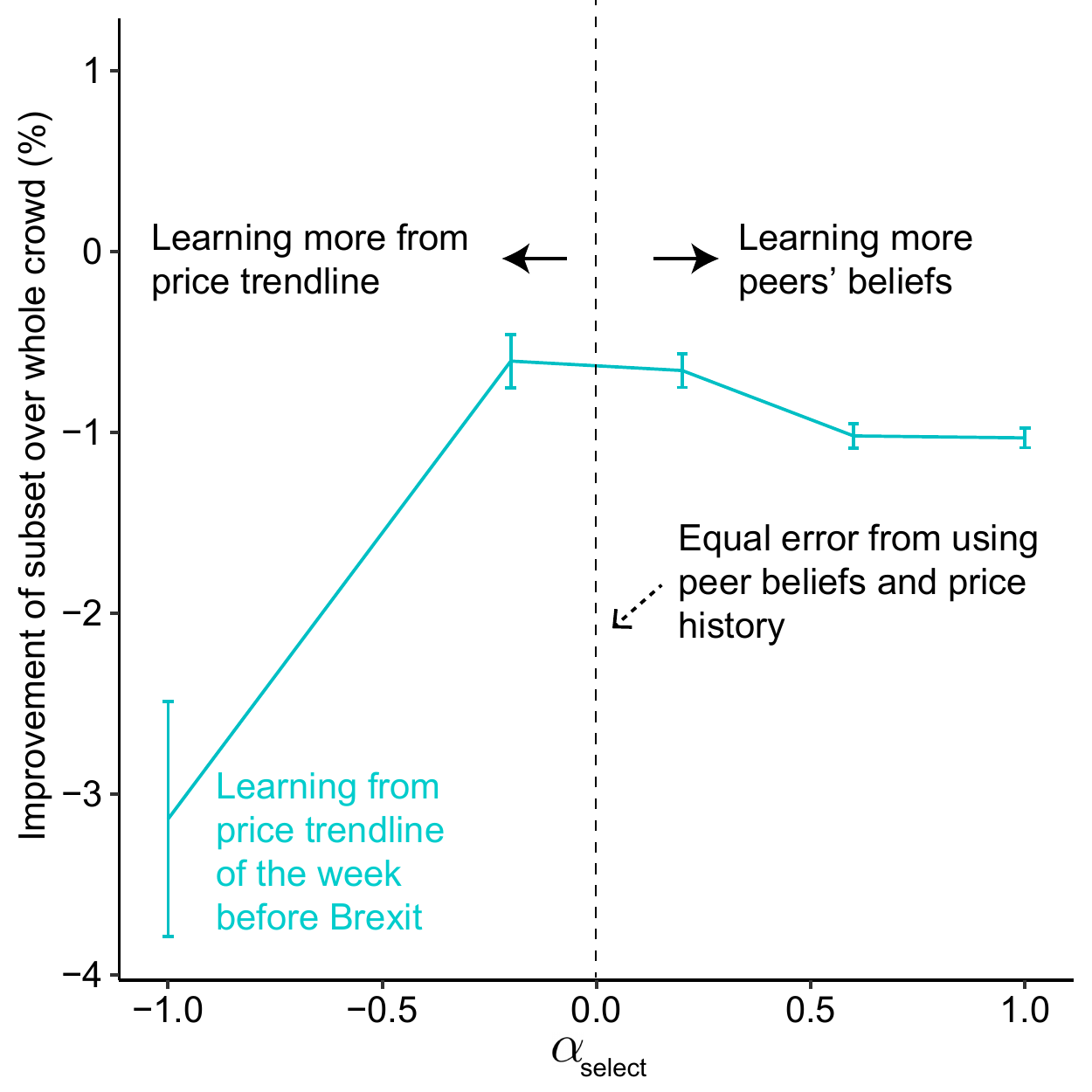}
        \caption{
        Improvement when selecting predictions based on how much more they were likely made using social information ($\alpha_{s} > 0$) vs. price history ($\alpha_{s} < 0$). 95\% Confidence intervals obtained through 100 bootstraps. When the price history is itself very uncertain, participants who learn from their peers do better than when they learned from the price history.}
        \label{fig:all_fanplots}
        \end{figure}%
        
        Again, we bin all $\alpha$'s from the prediction sets during this week and investigate the improvements of subsets of predictions compared to the whole crowd. We use a smaller number of bins due to the smaller number of predictions during the last week: 52 prediction sets in the last week compared to 284 during the open period of prediction that we previously used for predictions. 
        
        As can be seen in Fig. \ref{fig:all_fanplots}, as $\alpha$ decreases (i.e. we select predictions that were more likely updated using the price history instead of the social information), improvement in accuracy of subsets compared to the Wisdom of the Crowd (all predictions) severely decays down to -3.14\% (95\% confidence interval [2.49, 3.79]). Conversely, as subsets of predictions updated using the social histogram ($\alpha_{s} > 0$) are selected, the improvement in their accuracy is fairly stable (although negative).

        Note that although a smaller number of predictions were made during the last week before Brexit (52 prediction sets compared to 284 during the open period of prediction as discussed earlier), we have sufficient data to afford statistically significant results as shown by the 95\% confidence intervals of our findings.  The improvement values, confidence intervals, and their accompanying $\alpha_{s}$ are included in Table \ref{tab:improvement_brexit} in the supplementary. Unfortunately, given that such high market uncertainty only occurred during one round, we do not have enough data to produce a Pareto curve over multiple rounds.


        

        Our results suggest that although social learning generally leads to lower accuracy as shown in the Pareto curve of Fig. \ref{fig:Pareto}, during periods of high uncertainly, social learning leads to higher accuracy. This suggests that social learning could be leveraged by platform designers as a valuable tool that minimizes catastrophic performance.


    \subsection{Belief Update Models}
    \label{result:beliefupdate}
        Although our goal is not to search for the best model of individual belief update, we highlight our observations from fitting simple, interpretable, and theoretically-motivated models and distributions from prior work. It is important to note that \texttt{GaussianSocial} and \texttt{GaussianPrice} are both parameter-less models and did not require any parameter fitting, making their success in modeling belief update even more interesting. 

\begin{figure}
            \centering
            \includegraphics[width=.7\columnwidth]{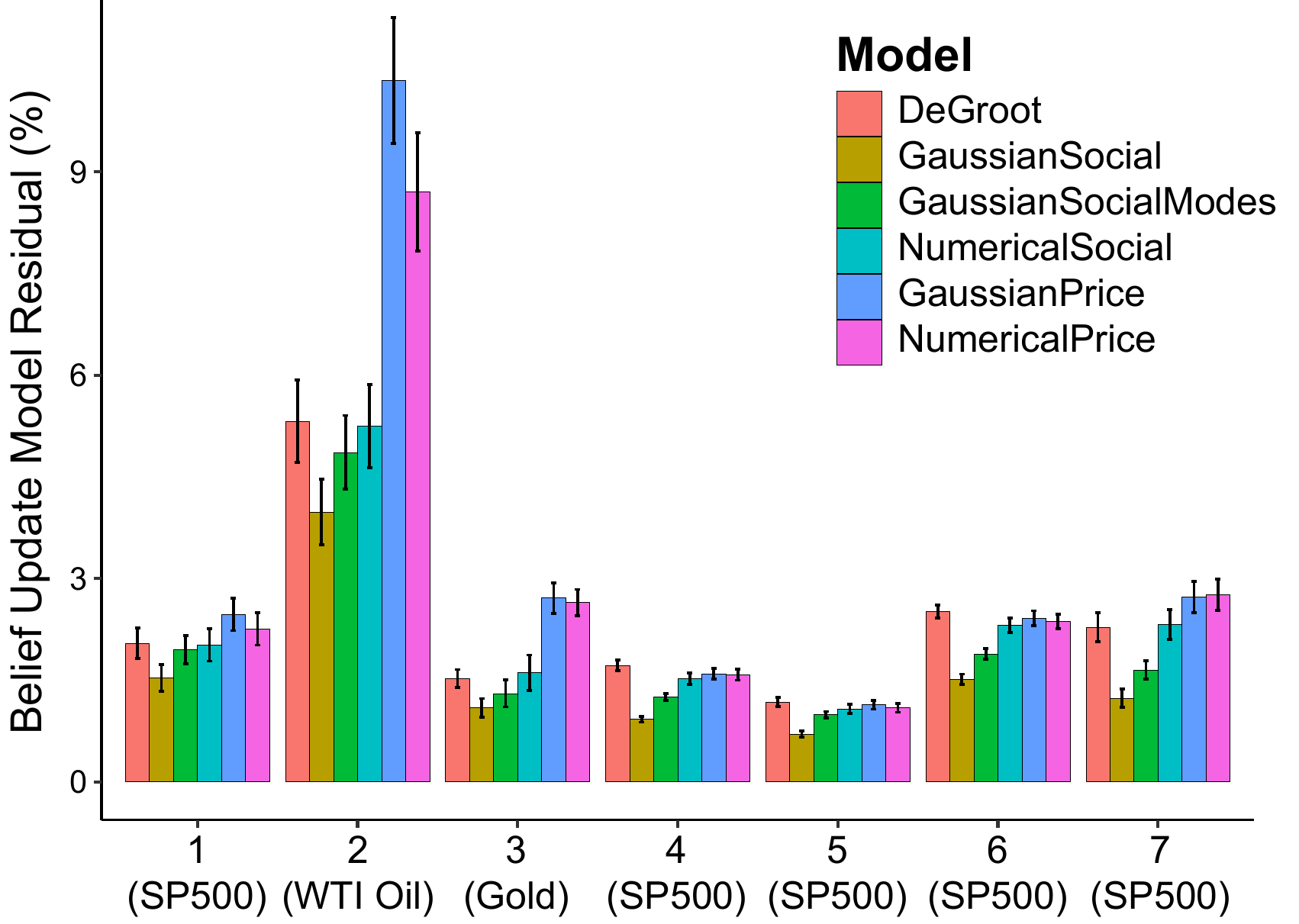}
            \caption{The y-axis shows the relative residual between \textit{modeled} belief update and \textit{actual} updated belief. Simple approximated models do better at modeling belief update than numerical models, and models using social histograms as likelihood perform better than models using the price history. Error bars represent 95\% CI.}
            \label{fig:simple_model_comparison}
        \end{figure}

        As can be seen in Fig.  \ref{fig:simple_model_comparison}, models that use social information for modeling the belief update of participants (\texttt{GaussianSocial},\texttt{Gaussian\-SocialModes}, \texttt{NumericalSocial}) perform better than models that use the price history (\texttt{GaussianPrice}, \texttt{NumericalPrice}). This suggests that our participants predominantly use social information instead of the price history to update their belief. 

        More specifically, \texttt{GaussianSocial}, our simple Gaussian model that assumes the data follows a single-mode Gaussian distribution, outperforms \texttt{GaussianSocialModes}, a model that measures when the social histogram is non-unimodal (which we estimate using the Hartigan's dip test of unimodality \cite{hartigan1985dip}) and uses the largest mode as the mean of the distribution in the same belief update procedure as \texttt{GaussianSocial}. This suggests that people assume the data they learn from to be unimodal even when it is non-unimodal, in line with prior work \cite{donnelly2006breast, nisbett1980human}. 

        Additionally, \texttt{GaussianSocial} outperforms the more precise numerical model \texttt{NumericalSocial} which makes no parametric assumption on the data distributions and uses a Monte Carlo procedure to estimate the posterior distribution. This suggests that people use simple heuristics when learning from their peers. However, when people are learning from the price history, the dominance of simpler models is not as clear as the performance of the simple \texttt{GaussianPrice} model is indistinguishable from that of the numerical model (\texttt{NumericalPrice}).
        
        \texttt{GaussianSocial} also outperforms the popular \texttt{DeGroot} model commonly used as a benchmark in the literature \cite{degroot1974reaching}, where an individual updates their belief as the weighted average belief of their peers. Here we set the weights (trust values) equal for all peers, as we have no data to estimate these weights, and therefore assume a uniform prior. It is interesting to note that \texttt{GaussianSocial} is equivalent to the \texttt{DeGroot} model when a participant's weight on their own prior belief is equal to the total of the weights of all other participants. This is in agreement with previous work showing that people put a disproportionately larger weight on their own prior belief \cite{dave2003confirmation,nickerson1998confirmation}. 
        
        Overall, the superiority of \texttt{GaussianSocial} in predicting belief update suggests that participants use a heuristic, uni-modal, and simple belief update procedure when updating their beliefs, and that they predominantly update their predictions using social information instead of price history. 

\section{Discussion}

    
    Our hypothesis was that a Pareto frontier exists between risk and accuracy --- i.e. that there is a trade-off between risk and accuracy of prediction --- and that social learning impacts this trade-off. Our results support this hypothesis: Fig. \ref{fig:Pareto} demonstrates that a Pareto frontier exists --- similarly to statistical \cite{james2013introduction, domingos2000unified, geman1992neural} and financial \cite{joyce1970uncertainty, modigliani1997risk, ghysels2005there} forecasting systems --- and that it is mediated by social learning. Additionally, our belief update models (Fig. \ref{fig:simple_model_comparison}) suggest that participants rely on social learning (and other heuristics) to update their belief. 
    
    
    \subsection{Design Implications}
    \label{discussion:design}
    
        Currently, crowdsourced prediction systems using a Wisdom of the Crowd approach focus on measuring and optimizing the average accuracy of their participants with little regard to the variance (risk) of the predictions. Practically, this means that platform designers and operators deploy a task to be predicted (e.g. predicting future prices) and end up with a single measure of performance for the task - the average accuracy of the crowd (or some similar measure of central tendency). 
        
        Our results suggest that the performance of such systems can be more comprehensively characterized by using both the average accuracy of the crowd and its variance. This is especially important because crowdsourced systems are being increasingly deployed for applications where uncertainty and risk can be quite harmful, such as in the medical information discovery domain \cite{lardon2015adverse, wang2016extraction, gottlieb2015ranking} where incorrect predictions are extremely costly. More generally, the modeling of risk supports more powerful and versatile applications of crowdsourced predictions such as hedging risks over portfolios of tasks which is standard in financial and statistical forecasting. 
        
        Additionally, given that the performance of a crowdsourced prediction system lies along a Pareto frontier, a practical question for designers is how to \textit{tune} the platform to reach a desired value of risk and accuracy. Our result that social learning can mediate the accuracy-risk trade-off provides a practical means to attain performances along this frontier. Practically, our results suggest that social learning within a crowdsourcing platform could be more purposefully leveraged to fit the task at hand: for example, platform designers could vary the social learning between users by incentivizing it to have lower risk --- especially during highly uncertain times, as our results from the Brexit prediction (Fig. \ref{fig:all_fanplots}) round showed. 
        
        Additionally our results provide evidence that there exists competing effects of exposure to social information versus non-social information on both accuracy and risk. Prior work had separately investigated exposure to social information \cite{Lorenz2011} or to non-social information \cite{hogarth1992order, payne1993adaptive}. By designing our experimental procedure such that people were freely able to learn from either social or non-social information and then estimating how much more of each source of information a person learned from, we are able to show that each type of learning causes opposite effects in terms of accuracy and risk: learning from the price history encourages higher accuracy, while learning from one's peers minimizes risk. This provides direct insights as to the design of crowdsourced prediction platforms as it indicates that there is an important balance between providing social and non-social information.

    \subsection{Heuristics and Biases}
    
    Our results also have implications for the literature on decision heuristics and biases \cite{nisbett1980human, tversky1974judgment}. Through the modeling of belief update, we observe that our subjects exhibit the attribute substitution heuristic of human decision-making \cite{Kahneman2014}, whereby a complicated problem is solved by approximating it with a simpler, less accurate model. We observe this heuristic as our participants simplify the data they are using to update their belief. This is evidenced by the fact that our participants' updated beliefs are better modeled by the \texttt{Gaussian\-Social} model (which assumes the data to be unimodal) than by the multi-modal belief update model \texttt{GaussianSocialModes}, indicating that our participants generally wrongly assume the data to be unimodal even when it is not (measured using the Hartigan's dip test of unimodality \cite{hartigan1985dip}). This is in line with previous studies that have shown that people wrongly assume data to be unimodal \cite{donnelly2006breast, nisbett1985perception, lindskog2013intuitive} due to the fact that using multi-modal data is cognitively costly \cite{hoffman2010costs}. Additional evidence of this substitution heuristic is from the fact that approximate models better predict the updated beliefs of participants than the more complicated (numerical) models: the \texttt{Gaussian\-Social} model outperforms the more precise Monte Carlo numerical models (as shown in Fig. \ref{fig:simple_model_comparison}). 
    
    Another decision heuristic that we observe is that people prefer to use social information rather than the underlying price history of an asset to update their belief as models which use social information (\texttt{GaussianSocial},\texttt{Gaussian\-SocialModes}, \texttt{NumericalSocial}) outperform models that use price history (\texttt{GaussianPrice}, \texttt{NumericalPrice}) as shown in Fig. \ref{fig:simple_model_comparison}. However, this collective preference for social learning comes at the price of lower accuracy (Fig.  \ref{fig:Pareto}). It is especially surprising that our participants preferred to use social information instead of prices to update their belief given that they were mid-career finance professionals with strong financial experience who should know that price information is generally better to predict future prices \cite{malkiel1970efficient, fama1965behavior}. Such behavior has been observed in prior work where even experts performing a familiar task demonstrate sub-optimal decision heuristics \cite{shanteau1988psychological, koehler2002calibration}, and often over-rely on social information \cite{foster1996strategic, posada2006learning}. However, instead of seeing such behavior as irrational, our results suggests that perhaps participants are preferentially aiming for lower risk instead of higher accuracy. This preference for social information especially pays off during the high uncertainty period before the Brexit vote.
    
    
    Therefore, our results support growing evidence that heuristics and biases are not merely \textit{defects} of human decision-making, but that perhaps they optimize for richer objectives or are optimized for more time- or data-constrained decision-making \cite{lakshminarayanan2011evolution,mallpress2015risk,kenrick2013rational,josef2016stability,cronqvist2014genetics,santos2015evolutionary,mishra2014decision}. For example, when individual decision-making is viewed within the lens of more realistic requirements such as limited time \cite{azuma2006review,cohen2008improving} or attention \cite{van2015information}, heuristics and biases---such as people assuming that the environment around them undergoes strong abrupt changes even when it is quite stable \cite{ryali2018demystifying, ma2015statistical}---act as priors that facilitate fast decision-making \cite{lubell2001cooperation,rand2016social}, and are quite helpful in practice.  

    \subsection{Future Work}
        Our work demonstrates that crowdsourced prediction platforms behave similarly to financial and statistical forecasting systems in that they exhibit an accuracy-risk Pareto frontier, and that this trade-off is mediated by social learning. This observation opens a number of interesting avenues for future work within the CSCW community. One interesting next step would be to investigate if different modalities of social communication and learning have have a similar accuracy-risk trade-off such as different types of discussions on forums \cite{krafft2018experimental} or the diversity of backgrounds of people interacting \cite{van2004work}. In our work, we restricted each round to have a static population of participants whose predictions were shared: an interesting direction for future work would be to embed participants in social networks given the importance and popularity of recent work on the effect of communication topologies \cite{adjodahNetES,almaatouq2020adaptive,becker2017network,barkoczi2016social} on group performance. Another interesting avenue for future work would be to utilize established metrics of risk aversion \cite{pratt1978risk} and investigate how subsetting predictions using these metrics affects collective accuracy and risk minimization. We also leave to future work the exploration of the large space of parametric and non-parametric models that best model people's belief update process. All these directions of future work pave the way for improving the design and deployment of crowdsourced prediction platforms.

    
    
    

    
\bibliographystyle{ACM-Reference-Format}
\bibliography{main}

\newpage
\onecolumn

\section*{Supplementary Material}

\section*{Derivation of Gaussian Model}
    
    We describe \texttt{GaussianSocial} here. \texttt{GaussianPrice} follows the same derivation, substituting the social histogram ${B_H}$ with the price history ${B_T}$. 
    
    Our notation follows that of \cite{kim2019bayesian}. We assume that people's estimate of the future price before information exposure, $B_{pre}$, is being sampled from an internal prior distribution \cite{vul2008measuring}, and that the sample we obtain is indicative of the mean of the prior distribution following the results of \cite{vul2014one}.

    We suppose that people think each asset has a true value, $V^*$, which people are trying to estimate to predict the future asset value, $V$ (the ground truth); that prior beliefs about $V^*$ follow a Normal (Gaussian) prior distribution, $V^* \sim Normal(\mu_{prior},\sigma_{prior})$; and that evidence about $V^*$ can be understood as being generated from a Normal distribution, $Normal(V^*,\sigma_{data})$.  In this case the posterior beliefs people have follows a simple form. Letting information content be defined as the inverse of the Normal distribution's variance $I = \frac{1}{\sigma}$, we have that
    
    \begin{equation}
        \mu_{posterior} = \frac{\mu_{prior} \cdot I_{prior} + \mu_{data} \cdot I_{data}}{I_{prior} + I_{data}}.
    \end{equation}

     Additionally, the social histogram is treated as representing the information content of data about $V^*$, then we have:
    
    \begin{equation}
        \mu_{posterior} = \frac{B_{pre} \cdot I_{prior} + \overline{{B_H}} \cdot I_{data}}{I_{prior} + I_{data}}.
    \end{equation}

    The \texttt{GaussianSocial} rule therefore can be viewed as reflecting an assumption of a Normal distribution as a mental model, and assuming private information and social information have the same information content ($I_{prior} = I_{data}$), which gives: 
    
    \begin{equation}
        \mu_{posterior} = \frac{B_{pre} + \overline{{B_H}}}{2}.
    \end{equation}

\section*{Performance over all rounds}

Here we report the values of the residual for each round for all models. We can observe that \texttt{GaussianSocial} does best. 
\begin{table}[H]
\centering
\resizebox{.9\textwidth}{!}{%
\begin{tabular}{r|l|l|l|l|l|l|l|}
\cline{2-8}
\multicolumn{1}{l|}{} & \multicolumn{7}{c|}{\textbf{ROUND}} \\ \hline
\multicolumn{1}{|c|}{\textbf{MODEL}} & \multicolumn{1}{c|}{\textbf{\begin{tabular}[c]{@{}c@{}}1 \\ (S\&P 500)\end{tabular}}} & \multicolumn{1}{c|}{\textbf{\begin{tabular}[c]{@{}c@{}}2\\ WTI Oil\end{tabular}}} & \multicolumn{1}{c|}{\textbf{\begin{tabular}[c]{@{}c@{}}3\\ Gold\end{tabular}}} & \multicolumn{1}{c|}{\textbf{\begin{tabular}[c]{@{}c@{}}7\\ (S\&P 500)\end{tabular}}} & \multicolumn{1}{c|}{\textbf{\begin{tabular}[c]{@{}c@{}}8\\ (S\&P 500)\end{tabular}}} & \multicolumn{1}{c|}{\textbf{\begin{tabular}[c]{@{}c@{}}9\\ (S\&P 500)\end{tabular}}} & \multicolumn{1}{c|}{\textbf{\begin{tabular}[c]{@{}c@{}}12\\ (S\&P 500)\end{tabular}}} \\ \hline
\multicolumn{1}{|r|}{\texttt{GaussianSocial}} & 1.53 (0.19) & 3.97 (0.48) & 1.08 (0.13) & 0.92 (0.04) & 0.70 (0.04) & 1.51 (0.07) & 1.23 (0.13) \\ \hline
\multicolumn{1}{|r|}{\texttt{GaussianSocialModes}} & 1.94 (0.20) & 4.85 (0.54) & 1.30 (0.19) & 1.24 (0.05) & 0.98 (0.04) & 1.88 (0.08) & 1.64 (0.13) \\ \hline
\multicolumn{1}{|r|}{\texttt{NumericalSocial}} & 2.01 (0.23) & 5.24 (0.61) & 1.60 (0.25) & 1.52 (0.08) & 1.07 (0.06) & 2.31 (0.10) & 2.31 (0.22) \\ \hline
\multicolumn{1}{|r|}{\texttt{NumericalPrice}} & 2.25 (0.23) & 8.70 (0.87) & 2.64 (0.19) & 1.57 (0.08) & 1.09 (0.06) & 2.36 (0.10) & 2.75 (0.23) \\ \hline
\multicolumn{1}{|r|}{\texttt{GaussianPrice}} & 2.46 (0.24) & 10.3 (0.92) & 2.70 (0.22) & 1.59 (0.07) & 1.13 (0.06) & 2.41 (0.10) & 2.72 (0.22) \\ \hline
\multicolumn{1}{|r|}{\texttt{DeGroot}} & 2.04 (0.22) & 5.32 (0.60) & 1.52 (0.13) & 1.71 (0.07) & 1.17 (0.06) & 2.51 (0.09) & 2.27 (0.21) \\ \hline
\end{tabular}%
}
\caption{Values of the residual for each round for all models. Numbers in parentheses show the 95\% error.}
\label{tab:my-table}
\end{table}

\section*{Table of Subsetting}

In this section, we report the improvement when selecting a subset of participants.
\begin{table}
\centering
\begin{tabular}{|l|c|l|}
\hline
\multicolumn{1}{|c|}{\textbf{$\alpha_{s}$}} & \textbf{\begin{tabular}[c]{@{}c@{}}Improvement (\%) \end{tabular}} & \multicolumn{1}{c|}{\textbf{95\% CI}} \\ \hline
-1.0 & 1.03 & 0.02 \\ \hline
-0.9 & 0.77 & 0.05 \\ \hline
-0.7 & 0.33 & 0.07 \\ \hline
-0.6 & 0.32 & 0.07 \\ \hline
-0.4 & 0.29 & 0.07 \\ \hline
-0.3 & 0.34 & 0.06 \\ \hline
-0.1 & -0.17 & 0.02 \\ \hline
0.0 & -0.48 & 0.06 \\ \hline
0.1 & 0.56 & 0.03 \\ \hline
0.3 & 0.38 & 0.03 \\ \hline
0.4 & 0.32 & 0.03 \\ \hline
0.6 & -0.27 & 0.08 \\ \hline
0.7 & -0.35 & 0.06 \\ \hline
0.9 & -0.67 & 0.04 \\ \hline
1.0 & -0.93 & 0.02 \\ \hline
\end{tabular}%
\caption{Improvements achieved by subsetting predictions via $\alpha_{s}$ for all rounds. Confidence intervals are calculated through 100 bootstraps.}
\label{tab:improvement_all}
\end{table}

\begin{table}
\centering
\begin{tabular}{|l|c|l|}
\hline
\multicolumn{1}{|c|}{\textbf{$\alpha_{s}$}} & \textbf{\begin{tabular}[c]{@{}c@{}}Improvement (\%)\end{tabular}} & \multicolumn{1}{c|}{\textbf{95\% CI}} \\ \hline
-1.0 & -3.14 & 0.65 \\ \hline
-0.2 & -0.61 & 0.15 \\ \hline
0.2 & -0.66 & 0.09 \\ \hline
0.6 & -1.02 & 0.07 \\ \hline
1.0 & -1.03 & 0.05 \\ \hline
\end{tabular}%
\caption{Improvements achieved by subsetting predictions via $\alpha_{s}$ only for predictions the week before Brexit. Confidence intervals are calculated through 100 bootstraps.}
\label{tab:improvement_brexit}
\end{table}

\begin{table}
\centering
\begin{tabular}{|c|c|c|}
\hline
\textbf{$\alpha_{s}$} & \textbf{Improvement (\%)} & \textbf{Standard Deviation} \\ \hline
-1.0 & 1.03 & 2.17 \\ \hline
-0.9 & 0.77 & 1.44 \\ \hline
-0.7 & 0.33 & 0.90 \\ \hline
-0.6 & 0.32 & 0.93 \\ \hline
-0.4 & 0.29 & 0.87 \\ \hline
-0.3 & 0.34 & 0.78 \\ \hline
-0.1 & -0.17 & 0.77 \\ \hline
0.0 & -0.48 & 0.62 \\ \hline
0.1 & 0.56 & 1.18 \\ \hline
0.3 & 0.38 & 1.21 \\ \hline
0.4 & 0.32 & 1.51 \\ \hline
0.6 & -0.27 & 1.18 \\ \hline
0.7 & -0.35 & 0.95 \\ \hline
0.9 & -0.67 & 0.70 \\ \hline
1.0 & -0.93 & 0.74 \\ \hline
\end{tabular}%
\caption{Improvement and bootstrapped Standard Deviation used in Pareto curve.}
\label{tab:my-table}
\end{table}

\end{document}